\documentclass[twocolumn,showpacs,amsmath,amssymb]{revtex4}
\usepackage{amssymb}

\usepackage{latexsym}
\usepackage{graphicx}
\usepackage{slashed}
\usepackage{pifont}
\usepackage{color}

\newcommand{\req}[1]{Eq.\,(\ref{#1})}

\newcommand{\beqn}{\begin{equation}}
\newcommand{\eeqn}{\end{equation}}

\begin{document}

\title{Pair Production from Asymmetric Head-on Laser Collisions}
\author{Lance Labun and Johann Rafelski}
\affiliation{Department of Physics, University of Arizona, Tucson, Arizona, 85721 USA \\
TH Division, Physics Department, CERN, CH-1211 Geneva 23, Switzerland}

\date{29 July, 2011} 

\begin{abstract} 
We evaluate particle production in highly asymmetric head-on collisions of  lasers pulses due to non-perturbative coherent action of many photons.
We obtain the yield of electron-positron pairs, which is controlled by the photon content of the weaker pulse, and show that the wavelength of the weaker pulse and the momentum asymmetry determine laboratory energy of the produced particles. 
\end{abstract}
%03.50.De Classical electromagnetism, Maxwell equations 
%11.15.-q Gauge field theories 
%11.15.Tk Other nonperturbative techniques 
%12.20.-m Quantum electrodynamics 
%12.20.Ds Specific calculations 
%12.20.Fv Experimental tests (for optical tests in quantum electrodynamics, see 42.50.Xa) 
%13.40.-f Electromagnetic processes and properties 
%42.50.Xa Optical tests of quantum theory 
%42.55.-f Lasers 
%52.27.Ep Electron-positron plasmas 
%52.38.Ph X-ray, γ-ray, and particle generation 
%52.38.Dx Laser light absorption in plasmas (collisional, parametric, etc.) 

\maketitle

\vspace*{-7.1cm}
\noindent CERN-TH-PH/2011-186\\
\vspace*{5.825cm}

\underline{\it Overview}. 
There is great interest in using lasers to produce beams of high energy particles.  We have previously shown that lasers can create electrons and positrons with high initial energies by {\it spontaneous} decay of the electromagnetic field~\cite{Labun:2011xt}, which requires extreme laser intensities to produce a significant number of particles.  Time dependence of the laser pulses can also {\it induce} pair production.  Circumstances where this process can add to the spontaneous process have already been reported~\cite{Schutzhold:2008pz,Dunne:2009gi}.  Here we show that with high energy pulsed lasers available today the induced process by itself can produce a large number of electron-positron pairs with ultra-high initial energies.

The coherence of the laser pulses means the center-of-momentum (CM) frame of a collision of laser pulses is defined by balancing the collective momentum of the photons in both pulses.  Therefore, the CM frame moves relative to the lab in the direction of propagation of the stronger pulse.  By choosing the boost $\gamma$ to the CM frame to Doppler-shift the frequency of the weaker laser field above threshold  $\hbar \tilde\omega > 2mc^2=1 \:{\rm MeV}$ (tildes mark quantities evaluated in the CM frame, and the background field provided by the other pulse can induce conversion of these MeV photons into electron-positron pairs.  

With eV-frequency laser pulses in the lab, a minimum boost $\gamma\sim 10^6$ is required to achieve threshold for electron-positrons pairs.  This implies we must consider in the laboratory frame (lab) a head-on collision of two laser pulses with momentum densities differing by a factor $\gamma^2\simeq 10^{12}$~\cite{Patent01}.   The energy of the pairs in the lab is necessarily $\gamma^{2}\:{\rm eV}$, i.e. TeV scale.  The appearance of the squared boost factor is analogous to the relativistic mirror effect: Going to the CM frame reveals the electron-positron pairs latent in the photons of the weak pulse, which can be reflected by the coherent field of the strong pulse.  We expect this effect is more robust than reflections of laser pulses from clouds of electrons, which suffer from incoherence of the mirror realization. 

Conversion of MeV photons to pairs in a background field is described by the established perturbative ${\cal O}(\alpha/\pi)$ expression for the total yield (summed over spins)~\cite{Itzykson:1980rh,Elze:2002bt}
\begin{align}
\label{Nexp} \Sigma_{(1)} =
  \frac{\alpha}{3}\!\int\!\frac{d^4k}{(2\pi)^4}\left(|\vec E(k)|^2-|\vec B(k)|^2\right) R_2(k^2), \\
\label{2body} R_{2}(k^2) \equiv
  \sqrt{1-\frac{4m^2}{k^2}}\left(1+\frac{2m^2}{k^2}\right)\Theta(k^2-4m^2)
\end{align}
Here, $\alpha\!=\!e^2/4\pi$ is the fine-structure constant, $\vec E(k)$, $\vec B(k)$ are the Fourier transforms of the electromagnetic fields, and $\Theta(x)$ is the unit step function.  $R_2(k^2)$ is the invariant 2-body phase space for fermions.  Pairs can be emitted only from the volume where the pulses overlap, because in a collision between a strong ({\bf s}-subscript) and weak ({\bf w}-subscript) linearly polarized pulse with aligned polarizations, 
%\beqn
$
\vec E^2-\vec B^2 = 
 2\vec E_{\rm s}\cdot \vec E_{\rm w}-2\vec B_{\rm s}\cdot \vec B_{\rm w}
$ and $\vec E\cdot \vec B\equiv 0$.

$\Sigma_{(1)}$ in \req{Nexp} being a linear function of  $|\vec E_{\rm s} \cdot \vec E_{\rm w}|$ shows that only one photon from each laser pulse contributes to each pair produced in the lowest order process evaluated in the CM frame.  We achieve a result non-perturbative in photon number in lab by being naturally compelled to work in a relativistic frame of reference where the dominant contribution comes from the term lowest order in $\alpha/\pi$.  In this sense, our technique is inspired by the infinite momentum frame formalism.  A strong field non-perturbative result becomes perturbative upon transforming into a frame of reference very near to the lightcone to identify participants in the interaction.  Indeed, the high energy of the produced pairs in the lab implies that $N_{\gamma}\sim 10^{12}$ photons have been absorbed from the strong pulse. 

Past work on pair production in head-on collisions of laser pulses has mainly addressed symmetric collisions~\cite{Brezin:1970xf,Avetissian:2002,Piazza:2004sv,Blaschke:2005hs,Bell:2008zzb}.  \req{Nexp} does not describe the yields in this case, requiring for the induced pair production process evaluation of non-perturbative processes which permit collective action of $\gamma^2$ photons combining into one pair~\cite{Fried:2001ga,Dunne:2010zz,Dumlu:2011rr}.  Only for highly asymmetric collisions, leading to the presence in the CM frame of above threshold energies, is the perturbative expression  \req{Nexp} physically meaningful.  Even though $\Sigma_{\!(1)}$ is a Lorentz invariant quantity and could be evaluated in any frame, application of this expression is restricted to physical circumstances which permit perturbative consideration.

%%%%%%%%%%%%%%%%%%%%%%%%%%%%%%%%%%%%%%%%%%%%%%%%%%%%%%%%%%%%
\underline{\it Collision dynamics in the CM frame}.
The Lorentz factor for the boost to the CM frame is $\gamma=(r\!+r^{-1} )/2,~\gamma v=(r\!-r^{-1})/2$, a function only of the ratio $r$ of pulse field strengths in the lab frame~\cite{Labun:2011xt}.  We adopt the convention that $r>1$, i.e. $r=|\vec E_{\rm s}|/|\vec E_{\rm w}|$. 

Note that the intensity of a pulse is the magnitude of its Poynting vector $|\vec S|=|\vec E \times \vec B|=\omega^2 a^2 m^2$ where $a=e|\vec A|/m$ the dimensionless laser amplitude of the laser pulse vector potential and $\omega$ is the pulse frequency.  The ratio of intensities therefore scales with $r^2$.  For the intensities to transform consistently, the laser amplitudes must transform as $\tilde a_{\rm s}^2 = a_{\rm s}^2r$ and $\tilde a_{\rm w}^2 = a_{\rm w}^2/r$.  This fact, counter-intuitive on first sight, arises from the $1/\sqrt{2\omega}$ normalization incorporated into $a$. 

The Doppler factor is $\gamma (1\pm v)=r^{\pm1}$.  For the weak pulse to be Doppler-shifted to an above-threshold frequency, $r$ must satisfy
\beqn\label{rdef}
\frac{|\vec S_{\rm s}|}{|\vec S_{\rm w}|} 
 = \frac{|\vec E_{\rm s}|^2}{|\vec E_{\rm w}|^2}
 = \frac{\omega^2_{\rm s} a^2_{\rm s}}{\omega^2_{\rm w} a^2_{\rm w}}
 \equiv r^2 >r_{\rm th}^2\equiv  \left(\frac{2m}{\omega_{\rm w}}\right)^{\!2}.
\eeqn

The Lorentz boost transforms the dynamics of the pulse collision by changing the unit proper time interval $d\tau$, which in the CM frame is $\gamma d\tau= dt$, and the relative lengths of the pulses.
The blue-shift of the wavelength of the weak pulse means its total length is contracted $\tilde L_{\rm w}=L_{\rm w}/r$.  The wavelength of the strong pulse is red-shifted to being quasi-constant, $\tilde\lambda_{\rm s} =r\lambda_{\rm s}$ and hence its total length is dilated $\tilde L_{\rm s}=r L_{\rm s}$.  
%In the CM frame the weak pulse is Lorentz-contracted into a pulse ultra-short compared to the Lorentz-stretched strong pulse.  
We shall see that the weak pulse can travel through the strong pulse for as long as it takes to convert most of its photons into pairs.

Viewing the collision in the CM frame and recalling that $\Sigma_{\!(1)}\propto |\vec E_{\rm s}\cdot\vec E_{\rm w}|$, we see that pair production occurs only in the compact region of the high frequency blue-shifted weak pulse as it traverses the quasi-constant red-shifted external field of the strong pulse.  The peak intensities of the two pulses are equalized in the CM frame.  The boosted intensity of the contracted weak pulse can be represented by its average, while the intensity of the dilated strong pulse rises and falls very slowly with the phase of the wave. 

\underline{\it Rate of Conversion into Pairs}. 
To determine the rate of conversion of coherent electromagnetic field into pairs, we consider the electromagnetic 4-momentum density, 
\beqn
p_{\rm e.m.}^{\mu} = T^{\mu\nu}u_{\nu},
\eeqn
being the only local conserved current available in the present context.  Here, $T^{\mu\nu}$ is the total electromagnetic energy-momentum tensor of the laser fields, and $u_{\nu}$ defines the hypersurface of the observer, becoming $u_{\nu}=(1,\vec 0)$ in the lab.  

Depletion of 4-momentum density of the strong and weak laser fields due to pair production is compensated by the 4-momentum density of created pairs, 
\beqn\label{dTmn}
\partial_{\mu}(p_{\rm e.m.}^{\mu}+p^{\mu}_{\rm pairs}) = 0.
\eeqn
We obtain the rate of 4-momentum transfer into created pairs by weighting the integrand of the pair yield \req{Nexp} with the 4-momentum $k^{\nu}$, 
\beqn\label{4momtransfer}
\Sigma_{(1)}^{\nu}=\frac{\alpha}{3}\!\int\!\frac{d^4k}{(2\pi)^4}\left(|\vec E(k)|^2-|\vec B(k)|^2\right)\:k^{\nu}R_2(k^2)
\eeqn
Then $u_{\nu}d\Sigma_{(1)}^{\nu}/d^4x$ describes the rate per unit time and space at which 4-momentum is extracted from the field in the frame characterized by $u_{\nu}$, and \req{dTmn} becomes
\beqn\label{transfereq}
\partial_{\mu}p_{\rm e.m.}^{\mu} =-u_{\nu}\frac{d\Sigma_{(1)}^{\nu}}{d^4x}. %=-\partial_{\mu}p_{\rm pairs}^{\mu}
\eeqn

Evaluating \req{transfereq} in the CM frame simplifies both sides to their 0-components:  For the left-side, the electromagnetic 3-momentum vanishes by virtue of having balanced the laser pulse momenta,
\beqn
\partial_{\mu}p_{\rm e.m.}^{\mu} = \tilde\partial_0\tilde{p}_{\rm e.m.}^0 \equiv d\varepsilon/d\tau
\eeqn 
where $\varepsilon$ is the energy density evaluated in the CM frame.  For the right side of \req{transfereq}, the ensemble of pairs has zero (average) net 3-momentum, $\langle \tilde{\vec u}\cdot \tilde{\vec{\Sigma}}_{(1)}\rangle = 0.$  Therefore,  in the CM frame \req{transfereq} is
\beqn\label{depsdt}
\frac{d\varepsilon}{d\tau} = -\tilde{u}_0\frac{d\tilde{\Sigma}^{0}_{(1)}}{d^4x}.
\eeqn

The upper limit for $\tilde{\Sigma}^{0}_{(1)}$ is obtained performing the 4-momentum integration replacing $R_2\to 1$, which is an over-estimate that nears the exact value for larger $r$.  The rate of energy transfer per unit time and space is  
\beqn\label{4momtransdens}
\frac{d\tilde{\Sigma}^{0}_{(1)}}{d^4x} < 
\frac{\alpha}{3}4|\vec E_{\rm s}\cdot \vec E_{\rm w}|(\tilde\omega_{\rm w}+\tilde\omega_{\rm s})
\eeqn 
(the coefficient $4$ arises from $|2\vec E_{\rm s}\cdot \vec E_{\rm w}-2\vec B_{\rm s}\cdot \vec B_{\rm w}|=4|\vec E_{\rm s}\cdot \vec E_{\rm w}|$).  This expression, being symmetric in the weak and strong pulse, shows that the energy of both pulses is depleted simultaneously.  In detail, the high energy and momentum of the produced pairs moving in the lab frame in the direction of the strong pulse requires that $r^2$ as much energy is derived from the strong pulse for each pair.  Because the strong pulse is  $r^2$ times as intense, $r$ as a function of time is unchanged by pair production and depends only on the rise and fall of the (background) strong pulse field in the CM frame, which is slow relative to the pair production dynamics.

The energy density $\varepsilon$ in the CM frame is obtained by writing the 4-vector momentum density as a unit vector multiplied by the magnitude, the invariant mass density $\sqrt{p^2_{\rm e.m.}}$, thus
\beqn\label{epseval}\begin{split}
\varepsilon &= \tilde{u}^0\sqrt{p_{\rm e.m.}^{\mu}p^{\rm e.m.}_{\mu}} \\ 
\sqrt{p_{\rm e.m.}^2} &= \sqrt{(\vec B^2-\vec E^2)^2/4+(\vec E\cdot\vec B)^2} 
= 2|\vec E_{\rm s}\cdot\vec E_{\rm w}|
\end{split}\eeqn

Plugging  Eqs.\eqref{4momtransdens} and \eqref{epseval} into the right side of \req{depsdt} gives a rate of attenuation of energy 
\beqn
\frac{d\varepsilon}{d\tau} \simeq -\frac{2\alpha}{3}(\tilde\omega_{\rm w}+\tilde\omega_{\rm s})\varepsilon,
\eeqn
proportional to the source.  For quasi-constant $r>r_{\rm th}$, we find an exponential attenuation law
\beqn\label{Ng}
\varepsilon(\tau) \simeq \varepsilon(0)e^{-\int^{\tau} \Lambda d\tau'}, 
 \quad \Lambda=\frac{2\alpha}{3}(\tilde\omega_{\rm w}+\tilde\omega_{\rm s}).
\eeqn
This expression is symmetric with respect to the strong and weak pulses.  However, the intensity asymmetry of the pulses results in the frequency asymmetry $\tilde \omega_{\rm s}/\tilde\omega_{\rm w}=\omega_{\rm s}/r^2\omega_{\rm w}$ in the CM frame, and the rate is primarily determined by the weak beam, $\tilde\omega_{\rm w}+\tilde\omega_{\rm s}\simeq\tilde\omega_{\rm w}$.  

Pair production sets in as the local $r$ nears $r_{\rm th}$ from below.  Once reaching $r_{\rm th}$ conversion of the pulses into pairs proceeds rapidly.  Achieving e.g. 3 times threshold energy in the CM frame $\tilde\omega_{\rm w}=3\:{\rm MeV}$ from a $2^{\rm nd}$ harmonic $\omega_{\rm w}=4.7\:{\rm eV}$ ($r=6.5\:10^5$), the rate $\Lambda$ corresponds to a lifespan of the system of $\tau=4\:10^{-20}\:{\rm s}$ in the CM frame and a time $t_0=14\:{\rm fs}$ in the lab frame.  This estimate shows that even within the fraction of the cycle of the strong pulse during which $r>r_{\rm th}$ a significant fraction of all energy in the pulses is converted into pairs.  Onset of pair production is made definite by having a pulse intensity front contrast ratio $> r$ in the strong pulse, so that its intensity switches within fraction of a cycle from $r$ below threshold to $r$ above threshold.  

Pair conversion in this context relies on the coherence of the laser pulses.  Though pairs are emitted as soon as the weak pulse touches the front of the strong pulse, photons from the entirety of each pulse are involved.  If the coherence length of the strong pulse is less than $t_0$, conversion to pairs will be considerablely less efficient, which favors short pulses for experimental implementation.  Focusing increases the intensity available in the strong pulse and hence the factor $r$, which must be large $\sim 10^6$ for the mechanism here to operate.  
 
%%%%%%%%%%%%%%%%%%%%%%%%%%%%%%%%%%%%%%%%%%%%%%%%%%%%%%%%%%%%%%%%%%%%%%%%%%%%

\underline{\it Colliding Gaussian Beams.} 
For a specific model case, we choose two colliding Gaussian beams with aligned polarizations.  
The lengths of the pulses $L_{\rm s},L_{\rm w}$ determine the longitudinal dimension and duration of the collision provided they are smaller than the respective Rayleigh lengths of the beams.  Both beams are near to their maximum focusing throughout the collision if the pulse lengths are less than one quarter the Rayleigh lengths.  To separate transverse focusing dynamics from the collision volume, we therefore require that for each pulse 
\beqn\label{Rayleighcond}
\pi(L_{\perp}/\lambda)_{\rm w,s}^2 > (L/\lambda)_{\rm w,s}
\eeqn
where $L_{\perp}$ is twice the beam waist.

The condition \req{Rayleighcond} permits us to consider the longitudinal dynamics as primarily controlling the field strengths in the collision volume.  We apply Gaussian envelopes to both pulses.  The strong pulse is so highly red-shifted that only the shape of the front matters.  With polarization vector $\vec\epsilon$ transverse to the direction of propagation, the vector potentials of the pulses are in position space
\beqn
\vec A_{\rm w,s}(x) = \vec\epsilon\:\frac{|\vec E_{\rm w,s}|}{\omega_{\rm w,s}}\:e^{-i\omega(t\pm z)}
e^{-2\vec x_{\perp}^2/L_{\perp}^2}e^{-2(t\pm z)^2/L_{\rm w,s}^2}
\eeqn
the upper sign applying to the weak pulse and the lower sign applying to the strong pulse.  
The Fourier transformed fields are then
\begin{widetext}
\vspace*{-0.2cm}
\begin{align}
\vec E^2(k)-\!\vec B^2(k)& = 
 4\pi^4|\vec E_{\rm s}\!\cdot\!\vec E_{\rm w}|L_{\!\perp}^4e^{-\frac{k_{\!\perp}^2L_{\!\perp}^2}{4}}\! 
   %\\ &\hspace*{0.3cm}\times\!
  \int\!dtdz\cos(\tilde\omega_{\rm s}(t\!-\!z))\cos(\tilde\omega_{\rm w}(t\!+\!z)) e^{-2\frac{(t+z)^2}{\tilde L_{\rm w}^2}}e^{-2\frac{(t-z)^2}{\tilde L_{\rm s}^2}}e^{i\omega t}e^{-ik_Lz}  \\
% \int\!dtdz\cos(\tilde\omega_s(t\!-\!z))\cos(\tilde\omega_w(t\!+\!z)) e^{-2(t+z)^2/\tilde L_{\!w}^2}e^{-2(t-z)^2/\tilde L_{\!s}^2}e^{i\omega t}e^{-ik_Lz}  \\
 &=\frac{\pi^6}{4}|\vec E_{\rm s}\!\cdot\!\vec E_{\rm w}|(L_{\!\perp}^2L_{\rm w}L_{\rm s})^2e^{-\frac{k_{\!\perp}^2L_{\!\perp}^2}{4}}
% &=\frac{\pi^4}{2}|\vec E_s\!\cdot\!\vec E_w|(L_{\!\perp}^2L_wL_s)^2e^{-k_{\!\perp}^2L_{\!\perp}^2/4} 
 %\\ &\hspace*{0.3cm}\times\!
\exp\Big(\!-\frac{\tilde L_{\rm w}^2}{32}(\omega-2\tilde\omega_{\rm w}-k_L)^2\Big) \sum_{\pm}\exp\Big(\!-\frac{\tilde L_{\rm s}^2}{8}(\omega-\omega_{\pm})^2\Big)
\label{FTfields}
\end{align}
\end{widetext}
Here, $\omega_{\pm}\equiv \tilde \omega_{\rm w}\pm\tilde\omega_{\rm s}$, and can be replaced $\omega_{\pm}\to\tilde\omega_{\rm w}$ to very good approximation, as noted below \req{Ng}.  

Because $L_{\perp}\gg m^{-1}$ and $L_{\perp}$ is unaffected by the transformation to the CM frame, the transverse momentum distribution is well-approximated by a $\delta$-function (numerical results are indistinguishable).  With $\tilde L_{\rm s}\gg m^{-1}$ the longitudinal shape of the highly red-shifted strong pulse similarly becomes unimportant.  

To evaluate the rate of particle production \req{Nexp}, we use \req{FTfields} in \req{Nexp}, and in the limit $L_{\perp},L_{\rm s}\to \infty$
\begin{align}\label{W1simple}
\frac{d\Sigma_{(1)}}{d^4x} &= 
 \frac{\alpha}{3} \frac{(2\pi)^{3/2}}{4}|\vec E_{\rm s}\!\cdot\!
    \vec E_{\rm w}|(L/\lambda)_{\rm w} I(r/r_{\rm th},\beta)\\
\begin{split}
I(r/r_{\rm th},\beta) &= 
\!\int_{-\infty}^{\infty}\!\!\!dk\: e^{-\beta (k+1)^2} \Theta(1-k^2-(r_{\rm th}/r)^{2}) \\ &\hspace*{0.4cm} \times
\sqrt{1-\frac{(r_{\rm th}/r)^{2}}{1-k^2}}\left(1+\frac{1}{2}\frac{(r_{\rm th}/r)^{2}}{1-k^2}\right)
\end{split}\raisetag{1.3cm}
\end{align}
in which $\beta=\pi^2(L/\lambda)_{\rm w}^2/8$.

\begin{figure}[tbh]
  \includegraphics[width=0.48\textwidth]{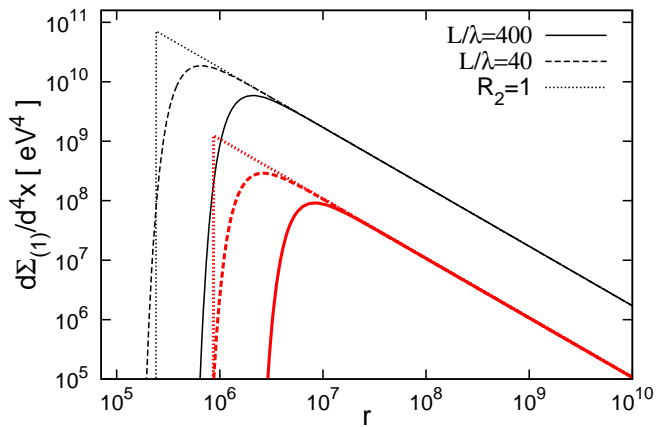}
\caption{The rate of converstion to electron-positron pairs as a function of the initial ratio of laser field strengths, $r$.  The magnitude of the strong laser pulse is fixed at $a=100$.  The key applies to both $\omega_{\rm w}=1.17\:{\rm eV}$ (lower set of thicker curves) and $\omega_{\rm w}=4.71\:{\rm eV}$ (higher, thinner curves), with the dotted curve showing $\Sigma_{(1)}$ computed with $R_2\to 1$. }
\label{fig:Nexp}       % Give a unique label
\end{figure}

We present in Figure~\ref{fig:Nexp} the numerically computed pair yield per unit of volume~\req{W1simple} for weak pulse lengths $L_{\rm w}=40\lambda_{\rm w}$ and $400\lambda_{\rm w}$ and for comparison the approximation $R_2=1$ used in \req{4momtransdens}.  The shorter pulse length exhibits its finite width by the onset of particle production below threshold, i.e. crossing the cutoff marked by the $R_2=1$ curve.  We see the rate is enhanced by using the $2^{\rm nd}$ harmonic of the weak pulse rather than equal strong and weak pulse frequencies $\omega_{\rm w}=\omega_{\rm s}=1.17\:{\rm eV}$.  Considering the typical interaction volume $\lambda_{\rm s}^3\simeq(1\:{\rm eV})^{-3}$, we can read the typical number of pairs per unit time directly from the vertical axis and expect $10^8-10^{10}$ pairs per $({\rm eV})^{-1}=4\:{\rm fs}$.

%%%%%%%%%%%%%%%%%%%%%%%%%%%%
%%%%%%%%%%%%%%%%%%%%%%%%%%%%

Higher order contributions to pair production depend only on field strength invariants.  Corrections to \req{Nexp} are therefore suppressed by powers of $\alpha |\vec E_{\rm s}\cdot \vec E_{\rm w}|/m^4_e=\alpha r^{-1}|\vec E_{\rm s}|^2/m^4$, a Lorentz invariant parameter made small by $r^{-1}$, i.e. the necessity of the weak pulse being much less intense than the strong pulse $|\vec E_{\rm w}|^2\lesssim 10^{-12}|\vec E_{\rm s}|^2$.  This condition further implies that $d\Sigma_{(1)}/d^4x\ll m^4$ is always satisfied, and \req{Nexp} is an accurate expression for the yield even when the strong pulse attains near-critical field strength $|\vec E_{\rm s}|\to m^2/e$ where spontanous pair production could occur.

%%%%%%%%%%%%%%%%%%%%%%%%%%%%%%%%%%%
\underline{\it Conclusions}.  We have presented a non-perturbative evaluation of induced electron-positron pair production from contemporary extreme-intensity pulsed lasers.  In asymmetric collisions of laser pulses, above-threshold energies are seen by going to the CM frame of the collision while strong external fields are generated by coherent action of a macroscopic number of photons.  The presence of the strong external potential inducing conversion into pairs is a feature found also in the context of the heavy ion collisions near to the Coulomb barrier~\cite{Reinhardt:1978cy,Reinhardt:1981zz}. 

Prior work has noted that induced pair production requires non-perturbative treatment to account for the many-photon process~\cite{Fried:2001ga,Dunne:2010zz,Dumlu:2011rr}.  We have solved this challenge with a technique inspired by light-cone formalism:  While the pulse collision can be studied in any frame, choosing the CM frame of highly asymmetric colliding pulses makes visible the dominant mechanism for the induced pair production---pair conversion of MeV-energy photons on an external field---and allows computation of the yield using available perturbative expressions~\cite{Itzykson:1980rh,Elze:2002bt}.  The boost into and out of the CM frame produces a relativistic mirror effect and pairs are seen in the lab with energies corresponding to the squared boost factor.  Electron-positron pairs produced by collisions of optical lasers appear in the lab with $(10^6)^2\:{\rm eV}={\rm TeV}$ energy.  A full evaluation of the spectrum requires more precise treatment of the back reaction process~\cite{Mihaila:2009ge} as well as the dynamics of the strong laser field.

Our result potentially offers a source of focused ultra-high energy elementary particle beams.  For preset ratio of intensities, the scale of particle energy can be controlled by modifying the frequency of the weak laser pulse.  The particle production cross-section scales with $m^{-2}$ as for other elementary processes, leading for example to the production of 1 muon-anti-muon pair per 40,000 electron-positron pairs when threshold for muon production is reached.  Although the threshold intensity ratio for muon production $r_{\rm th,\mu}^2\!\sim\!(2\:10^8)^2$ appears a technological challenge, the reward is production of muon-anti-muon pairs of $40\:{\rm PeV}$ energy in the lab, a previously unreachable energy.  

%\vskip 0.2cm
{\it Acknowledgments.}  This work was supported by a grant from the U.S. Department of Energy, DE-FG02-04ER41318.  The authors thank CERN-PH-TH for hospitality and G. Altarelli for valuable comments.

\end{document}